\documentstyle[preprint,aps,floats,psfig]{revtex}
\tighten
\begin{document}

\title{Stewart-Lyth inverse problem}

\author{Eloy Ay\'on-Beato$^1$\thanks{E-mail: ayon@fis.cinvestav.mx}, 
Alberto Garc\'{\i}a$^1$\thanks{E-mail: aagarcia@fis.cinvestav.mx}, 
Ricardo Mansilla$^2$\thanks{E-mail: mansy@servidor.unam.mx}, 
C\'esar A.\ Terrero-Escalante$^1$\thanks{E-mail: cterrero@fis.cinvestav.mx}}

\address{
$^1$Departamento~de~F\'{\i}sica,~CINVESTAV-IPN,~Apdo.~Postal~14-740,
~07000,~M\'exico~D.F.,~M\'exico.\\
$^2$Departamento~de~F\'{\i}sica~Te\'orica,~IF-UNAM,~Apdo.~Postal~20-364,
~01000,~M\'exico~D.F.,~M\'exico.}
\maketitle

\begin{abstract}
In this paper the Stewart-Lyth inverse problem is introduced. It
consists of solving two nonlinear differential equations for the
first slow-roll parameter and finding the inflaton potential. The
equations are derived from the Stewart-Lyth equations for the
scalar and tensorial perturbations produced during the inflationary
period. The geometry of the phase planes transverse to the
trajectories is analyzed, and conclusions about the possible
behavior for general solutions are drawn. 
\end{abstract}

\section{Introduction}

The analysis of observations of the cosmic microwave background radiation,
recently reported by the experiments Boomerang and Maxima-1
\cite{observ,observ2},
 confirmed that, at present-day, inflation \cite{Guth,Linde1}
still remains as the favourite model for the origin of structure in
the Universe. The main reason is that the currently obtained
observational data on structures in the Universe are interpreted in
a natural way within the framework of the Gaussian adiabatic and
nearly scale-invariant density perturbations that the usual models
produce. On the other hand, it is one of the simplest paradigms
within which rigorous theoretical predictions can be achieved.

However, the inflationary paradigm is quite a broad one, and there
are several equally satisfactory implementations of the
inflationary idea. The simplest scenario arises when the dynamics
of inflation (both classical and quantum) are dominated by a single
scalar field (inflaton) evolving in a nearly flat potential. Even
in this case, there are a large number of candidates for the
potential. For this scenario it is well established
\cite{Lid2,Lid3} that, to a good approximation, the scalar and
tensor perturbations will take on a power-law form, with the
tensor ones giving a subdominant (and almost negligible)
contribution. A way of discriminating between candidates for
inflationary potential is the comparison between observed and
predicted perturbations spectra. However, the present level of
accuracy of the observations is well below the accuracy of the
predictions for most of the models \cite{Lid3a}, allowing a 
large number of them to remain as candidates.

Another way of accomplishing the task of choosing a proper
potential is to reconstruct it from the observed spectra \cite
{Lid3a}, but again the low accuracy of measurements as well as the
fact that the reconstruction is based on a Taylor expansion of the
potential (with higher order derivatives given in dependence of the
amount and quality of observational data), implies a large
uncertainty in the outcomes of the standard reconstruction
procedure, from the point of view of the potential uniqueness \cite
{Lid3a}. Anticipating the near-future launch of satellites capable of
measuring microwave background anisotropies to an accuracy of a few
percent or better, across a wide range of angular scales
\cite{Page}, an assessment of the accuracy of predictions of the
anisotropies for given cosmological models has been stressed.
Grivell and Liddle \cite{Lid4} have confirmed that for most models
of inflation, the Stewart and Lyth analytic calculations \cite
{StLy} should give extremely accurate predictions.

Lyth and Stewart \cite{LySt} began with the precise calculation for
power-law inflation \cite{Lucchin} and then they went on to use
this exact result to analytically compute the next-order
slow-roll correction to the standard formula \cite{StLy}. At this
level of approximation, the Stewart and Lyth equations for the
spectral indices \cite{StLy} can be rewritten as nonlinear
differential equations in terms of the first slow-roll parameter
$\epsilon$ \cite{SchMi,Benitez}. Hence, a third option for the
determination of the inflationary potential, that we shall discuss
in this paper, is to use observational information about density
perturbations and gravitational waves spectra as input in these
differential equations, to solve for $\epsilon$ and to find
the corresponding inflationary potential. We call this procedure
the Stewart-Lyth inverse problem.

In this work, after introducing the Stewart-Lyth inverse problem,
we test the feasibility of the method by finding the proper
potential for the well-known scenario of power-law inflation.
Nevertheless, the involved system of differential equations
composed by a second order and a first order equations, have proved
to be very difficult to solve for general functional forms of the
perturbations spectra. Taking this into account, in this paper we
restrict our analysis to a qualitative study of the dynamics
described by these equations. To draw conclusions about the
behavior of the general solutions, the equations are studied using
the spectral indices as parameters. For the corresponding reduced
equations, we analyze the phase space for the second order equation
which presents a singularity that strongly determines the flow
geometry. Not cyclical orbits are found. The reduced first order
equation is solved for any value of the tensorial spectral index
and the possible behaviors are analyzed with all of the solutions
being monotonic. Except in the case of null tensorial index, the
singularity is also observed. In general, for solutions of the
Stewart-Lyth inverse problem with smoothly and slowly changing
spectral indices, periodic, quasiperiodic, or chaotic general
solutions should not be expected. The theoretical analysis  as well
as observations suggest that power-law solutions are still valid.

In the next section we briefly describe the theoretical frame for the
Stewart-Lyth calculations and present their originally algebraic
equations in terms of the spectral indices. In Sec.~\ref{sec:InvP}
we introduce the Stewart-Lyth inverse problem.
Section ~\ref{sec:phase} is devoted to the qualitative analysis of the
phase-spaces of the reduced second order equation for the first
slow-roll parameter. The analysis of the solutions for the reduced
first order equation is presented in Sec.~\ref{sec:first}. We
summarize the main results obtained in Sec.~\ref{sec:conclu}.

\section{The Stewart-Lyth equations}

\label{sec:SLeq}

\subsection{The single scalar field scenario}

The theoretical frame for the Stewart-Lyth calculation is the flat
Friedmann-Robertson-Walker universe containing a single scalar field
equivalent to a perfect fluid with equations of motions given by
\begin{eqnarray}
H^2 &=& \frac{\kappa}{3}\left(\frac{\dot{\phi}^2}{2} + V(\phi)\right),
\label{Friedmann} \\
\ddot{\phi} &+& 3H\dot{\phi} = -V^\prime(\phi) \, ,\label{mass}
\end{eqnarray}
where $\phi$ is the inflaton, $V(\phi)$ is the inflationary
potential, $H=\dot{a}/a$ is the Hubble parameter, $a$ is the scale
factor, dot and prime stand for derivatives with respect to cosmic
time and $\phi$ respectively, $\kappa = 8\pi/m_{\rm Pl}^2$ is the
Einstein constant and $m_{\rm Pl}$ is the Planck mass.

In this framework, the first three slow-roll parameters were
respectively defined in Ref.~\cite{Lid5}, we shall write them here as in Ref.
~\cite{Lid3a}:
\begin{eqnarray}
\epsilon(\phi) &\equiv& 3\frac{\dot{\phi}^2}{2}\left[\frac{\dot{\phi}^2}{2}
+ V(\phi)\right]^{-1} = \frac{2}{\kappa}\left[\frac{H^\prime}{H}\right]^2,
\label{SRP1} \\
\eta(\phi) &\equiv& -\frac{\ddot{\phi}}{H\dot{\phi}} = \epsilon -\frac{
\epsilon^\prime}{\sqrt{2\kappa\epsilon}}\, ,\label{SRP2} \\
\xi(\phi) &\equiv& \left[\epsilon\eta - \left(\frac{2}{\kappa}
\epsilon\right)^{\frac{1}{2}}\eta^\prime\right]^{\frac{1}{2}}.\label{SRP3}
\end{eqnarray}

Up to a constant, the first slow-roll parameter (\ref{SRP1}) is a
measure of the relative contribution of the kinetic energy to the
total energy of the field. By definition $\epsilon\geq0$, and
because inflation can be defined as $\ddot{a}>0$, for inflation to
proceed: $\epsilon<1$.

\subsection{Power-law inflation}

Few models of inflation allow to exactly calculate scalar and
tensorial perturbations. One of them is the power-law model
\cite{Lucchin}, a particular scenario of inflation, where
\begin{eqnarray}
a(t) &\propto& t^p \, , \label{PL}\\
H(\phi) &\propto& \exp\left(-\sqrt{\frac{\kappa}{2p}}\,\phi\right),
\label{HPL} \\
V(\phi) &\propto& \exp\left(-\sqrt{\frac{2\kappa}{p}}\,\phi\right),
\label{VPL}
\end{eqnarray}
with $p$ being a positive constant. It follows from Eqs.~(\ref{SRP1}),
(\ref{SRP2}), and (\ref{SRP3}) that in this case the slow-roll
parameters are constant and equal each other.

\subsection{The indices equations}

Exact expressions for the asymptotic scalar (density perturbations) and
tensorial (gravitational waves) power spectra for the case of power-law
inflation was correspondingly derived by Lyth and Stewart \cite{LySt} and
Abbot and Wise \cite{Abbot}.

Assuming that the deviation of the higher slow-roll parameters
from $\epsilon $ is small (power-law approximation) and that
$\epsilon$ is small with respect to unity (slow-roll
approximation), Stewart and Lyth \cite {StLy} derived
next-to-leading order expressions for both spectra. In terms of
the spectral indices, these expressions are
\begin{eqnarray}
1-n_s(k) &\simeq& 4\epsilon - 2\eta + 8(C+1)\epsilon^2 - (10C+6)\epsilon\eta
+ 2C\xi^2, \label{StLythNs} \\
n_T(k) &\simeq& -2\epsilon\left[1 + (2C+3)\epsilon - 2(C+1)\eta \right],
\label{StLythNt}
\end{eqnarray}
where the notation is that of \cite{Lid3a}; $n_s(k)$ and $n_T(k)$
are the scalar and tensorial spectral indices respectively, $k$ is
the wave number of the comoving scale, and $C \approx -0.73$ is a
constant related to the Euler constant originated in the expansion
of Gamma function. The symbol $\simeq$ is used to indicate that
these equations were obtained using the power-law and slow-roll
approximations. Hereafter we shall use the $=$ sign in our
calculations, but the meaning of approximation should be added
whenever it applies.

For a giving expression of the scale factor, the Hubble parameter
and the potential are determined, and then by substituting
definitions (\ref{SRP1}), (\ref{SRP2}), and (\ref{SRP3}) in the
Stewart-Lyth equations (\ref{StLythNs}) and (\ref{StLythNt}), the
scale-dependent spectral indices are obtained. For instance,
giving Eq.~(\ref{HPL}) and substituting in Eqs.~(\ref{SRP1}), (\ref {SRP2}),
and (\ref{SRP3}), one obtains $\epsilon=\eta=\xi=1/p$, which in
turn substituted in Eqs.~(\ref{StLythNs}) and (\ref{StLythNt}) yield
$n_s-1=n_T= {\rm const}$. This is an alternative definition for
power-law inflation. We should note that the spectral indices are
directly related with the power spectra (for details see Ref.~\cite{Lid3a}).

\section{Formulation of the Stewart-Lyth inverse problem}

\label{sec:InvP}

In this section we shall formulate a method for finding the
inflaton potential using observational information on the spectral
indices. Denoting $T\equiv\dot{\phi}^{2}/2$ and using definitions
(\ref{SRP1}), (\ref{SRP2}), (\ref{SRP3}), together with Eqs.
~(\ref{Friedmann}) and (\ref {mass}) then, in a straightforward
manner, we obtain
\begin{eqnarray}
\epsilon &=&3\frac{T}{T+V}=\frac{\kappa T}{H^{2}}\, ,\label{SRP1T}\\
\eta &=&\kappa \frac{dT}{dH^{2}}\, ,\label{SRP2T} \\
\xi ^{2} &=&\kappa \epsilon \frac{dT}{dH^{2}}
+2\kappa \epsilon H^{2}\frac{
d^{2}T}{d(H^{2})^{2}}\, .\label{SRP3T}
\end{eqnarray}
Defining $\tau \equiv \ln \,H^{2}$, $\delta (k)\equiv n_{T}(k)/2$
and $\Delta (k)\equiv [n_{s}(k)-1]/2$, and substituting Eqs.
~(\ref{SRP1T}), (\ref {SRP2T}), and (\ref{SRP3T}) in Eqs.
~(\ref{StLythNs}) and (\ref{StLythNt}), the indices equations in
terms of the first slow-roll parameter $\epsilon $ and its
derivatives with respect to $\tau $ ($\hat{\epsilon}\equiv
d\epsilon
/d\tau $ and $\hat{\hat{\epsilon}}\equiv d^{2}\epsilon /d\tau ^{2}$) become
\begin{eqnarray}
2C\epsilon \hat{\hat{\epsilon}}-(2C+3)\epsilon \hat{\epsilon}-\hat{\epsilon}
+\epsilon ^{2}+\epsilon +\Delta &=&0\,, \label{MSch1} \\
2(C+1)\epsilon \hat{\epsilon}-\epsilon ^{2}-\epsilon -\delta &=&0\,.
\label{MSch2}
\end{eqnarray}
The second order nonlinear differential equation (\ref{MSch1})
was first introduced in Ref.~\cite{SchMi}, while the first order
equation (\ref{MSch2}) was derived first in Ref.~\cite{Benitez}. Notice
that Eqs.~(\ref{MSch1}) and (\ref{MSch2}) are just alternative
representations of the Stewart-Lyth Eqs. (\ref{StLythNs}) and
(\ref{StLythNt}). The approach of Ref.~\cite{SchMi} for the potential
reconstruction is incomplete because is restricted to the case
$\Delta(k)={\rm const}$, and, on the other hand, the equation for
the tensorial spectral index (\ref{MSch2}) is considered only in
the trivial case of power-law inflation \cite {Benitez}. As we
shall see, Eq.~(\ref{MSch2}) imposes rigorous constrains
upon the set of solutions of Eq.~(\ref{MSch1}).

By the Stewart-Lyth inverse problem we mean the problem
consisting in solving Eqs.~(\ref{MSch1}) and (\ref{MSch2}) for
$\epsilon$, given expressions for the spectral indices, and finding
the corresponding inflaton potential using the definitions of the
first slow-roll parameter (\ref{SRP1}) and (\ref{SRP1T}). We
prefer this denomination instead of ``reconstruction'' because we
will use explicit functional forms of the spectra as input in the
problem and correspondingly we will obtain functional forms for the
potentials rather than pieces of information about them.

As was already mentioned, having an expression for $\epsilon(\tau)$ the
corresponding potential as a function of $\tau$ can be obtained from Eq.
~(\ref{SRP1T}):
\begin{equation}
V(\tau)= \frac{1}{\kappa}\left[3-\epsilon(\tau)\right]\exp(\tau)\,.
\label{PotentialT}
\end{equation}

On the other hand, taking into account that $\dot{\phi}=\dot{\tau} \hat{\phi}
$ and using Eq.~(\ref{SRP1}), the scalar field as a function of $\tau$ is given
by
\begin{equation}
\phi(\tau)= -\frac{1}{\sqrt{2\kappa}} \int\frac{d\tau}{\sqrt{\epsilon(\tau)}}
+ \phi_0\, ,  \label{PhiT}
\end{equation}
where $\phi_0$ is an integration constant.

Finally, the inflationary potential as parametric function of the inflaton
can be given:
\begin{equation}
V(\phi)= \left\{
\begin{array}{cc}
\phi(\tau)\, , &  \\
V(\tau)\, . &
\end{array}
\right.  \label{FVphi}
\end{equation}
The above expressions are similar to those used in Ref.~\cite{SchMi} but
we would like to stress that the functional form for $\epsilon$ in
expressions (\ref {PotentialT}) and (\ref{PhiT}) must be solution
of both Eqs.~(\ref{MSch1}) and (\ref{MSch2}).

As a simple example of a resolution of the Stewart-Lyth inverse
problem, let us analyze the case of power-law inflation, where the
spectral indices are constants. Substituting Eq.~(\ref{MSch2}) in
Eq.~(\ref{MSch1}) with constant $\Delta$ and $\delta$, an algebraic
equation for $\epsilon$ is obtained. Since the coefficients of this
equation are scale-independent, all the solutions for this
algebraic equation are just of the form: $\epsilon=1/p$ with
$p={\rm const}$. Substituting this expression for $\epsilon$ in
Eq.~(\ref{PhiT}) and after some algebra, we obtain for the Hubble
parameter
\begin{equation}  \label{HPR}
H(\phi)= \exp\left(-\sqrt{\frac{\kappa}{2p}}\,(\phi-\phi_0)\right).
\end{equation}

We can see that, up to a constant, Eq.~(\ref{HPR}) is equivalent to (\ref{HPL}).
Now, substituting $\epsilon=1/p$ and $H^2$ in Eq.~(\ref{PotentialT}), 
we obtain,
\begin{equation}
V(\phi)= \frac{3p-1}{\kappa
p}\exp\left(-\sqrt{\frac{2\kappa}{p}}\,(\phi-\phi_0)\right),\label{VPR}
\end{equation}
in complete correspondence with Eq.~(\ref{VPL}).

On the relevance of taking into account the first-order equation,
we want to remark that, although Eq.~(\ref{MSch1}) has a large
number of solutions for a constant scalar index, once Eq.
~(\ref{MSch2}) with $\delta={\rm const}$ was used as a first
integral of Eq.~(\ref{MSch1}), the unique remaining solution is just
the potential given by Eq.~(\ref{VPR}).

Even for a scale-independent scalar spectrum, Eq.~(\ref{MSch1}), 
when it is taken with no regards to Eq.~(\ref{MSch2}), has
proved to be very difficult to be analytically solved; solutions
have been found only for a few fixed values of $\Delta$ (see
Refs.~\cite{Benitez,Alberto} and references therein). This obstacle is
even harder to overcome when an scale-dependent scalar spectrum is
assumed; even for the first order Eq. (\ref{MSch2}) it is hard
to find an explicit solution for $\epsilon$ for any value of
$\delta ={\rm const}$, and even more difficult for $\delta
=\delta[\tau (k)]$.

In the present work, we will focus in the dynamical aspect of the
Stewart-Lyth inverse problem, i.e., in the dynamics of the first
slow-roll parameter determined by Eqs.~(\ref{MSch1}) and
(\ref{MSch2}). The problem of integrating for the inflationary
potential is left to be done in a near future.

\section{Phase spaces analysis of the reduced second order equation}
\label{sec:phase}

For an understanding of the dynamics behind Eqs.~(\ref{MSch1})
and (\ref {MSch2}), the best approach seems to be a qualitative
analysis of the corresponding phase-spaces. Not yet having an explicit 
expression for $\Delta(\tau)$ and
$\delta(\tau)$, we use the following approach in order to draw
conclusions about the dynamics described by Eqs.~(\ref{MSch1}) and
(\ref{MSch2}): if we consider $\Delta(\tau)$ and $\delta(\tau)$ as
the forcing element in these equations, then we can assume the
dynamics to be characterized by one more dimension. The $(\epsilon
,\hat{\epsilon})$ planes [$(\tau,\epsilon)$ for the first order
equation] corresponding to the different values of this new
coordinate are transverse to the trajectories given by these
equations. Having the phase-portraits on the planes
$(\epsilon,\hat{\epsilon })$ for any value of $\Delta={\rm const}$
in Eq.~(\ref{MSch1}) and the solutions $(\tau,\epsilon)$ for
any $\delta={\rm const}$ in Eq.~(\ref{MSch2}), the geometry
of the surfaces along which the real trajectories spread out could
be outlined, assuming slow variation for $\Delta(\tau)$ and
$\delta(\tau)$.

Hereafter, we will refer to Eqs.~(\ref{MSch1}) and
(\ref{MSch2}) with constant $\Delta$ and $\delta$ as the 
reduced equations of the Stewart-Lyth inverse problem, and the
space for solutions depending on $\tau $ as the extended
phase-space.

Solutions for the reduced first order equation (\ref{MSch2}) will
be studied in the next section. Let us now proceed with the
analysis of the dynamics given by Eq.~(\ref{MSch1}). This equation can
be rewritten as
\begin{equation}
\hat{\hat{\epsilon }}-\frac{1}{2C}\left( 2C+3 +\frac{1}{\epsilon} \right)
\hat{\epsilon}+\frac{1}{2C}\left( \epsilon +1+\frac{\Delta}{ \epsilon }
\right) =0\, .\label{TrEq}
\end{equation}

With the change of variables
\begin{equation}
\begin{array}{c}
x_{1}=\epsilon\, , \\
x_{2}=\hat{\epsilon}\, ,
\end{array}
\label{VarCh}
\end{equation}
we obtain the system
\begin{equation}
\begin{array}{l}
\hat{x}_{1}\equiv F\left(x_{1},x_{2}\right)=x_{2} \,, \\
\hat{x}_{2}\equiv G\left(x_{1},x_{2}\right)= -\frac{1}{2C}\left( x_{1}+1+%
\frac{\Delta }{x_{1}}\right) +\frac{1}{2C}\left( 2C+3+\frac{1}{x_{1}}\right)
x_{2} \,.
\end{array}
\label{DynSys}
\end{equation}
For $\Delta={\rm const}$, the condition for existence of solution
\cite{Elgoltz}, i.e., the continuity condition for the vector
field, holds at every point $(x_1,x_2)\in {\mathcal P}\equiv{\rm
{I\!R}}^2\backslash
\left\{ \left(0, x_2\right), \forall x_2 \right\}$. On the other hand,
\begin{equation}
\left| G(x_1,x_2^{(1)})-G(x_1,x_2^{(2)})\right| = \frac{1}{2\left| C\right|}%
\left|(2C+3) +\frac{1}{x_1}\right|\left| x_2^{(1)}-x_2^{(2)}\right|,
\label{Lipschitz}
\end{equation}
where the upper indices denote any two different values of $x_2$.
Then, the uniqueness condition, i.e., Lipschitz condition, holds
for the same set ${\mathcal P}$. Therefore, unique solution for
the equations system (\ref{DynSys}) certainly exists at any point in 
${\mathcal P}$. In
a similar fashion, the differentiability of solution with respect
to initial conditions and parameters of the system is also
satisfied. Hence, we can use the results of qualitative theory of
dynamical systems in the plane for the study of system (\ref{DynSys})
\cite{GH}.

The set of singular points of system (\ref{DynSys}) is the union of the set of
fixed points ($F=G=0$) and the $x_2$ axis. We should note that around point
$p_s=(0,\Delta)$ the field is well defined along directions for which $x_2
\rightarrow \Delta$ faster than $x_1 \rightarrow 0$. This case will be
analyzed in detail in Sec.~\ref{Point_ps}.

\subsection{Fixed points}

The fixed points of this system are the set $\{\left(x_{1},x_{2}\right)\}$
such that
\begin{eqnarray}
F\left(x_{1},x_{2}\right) &=&0\, ,  \nonumber \\
G\left(x_{1},x_{2}\right) &=&0\, .  \nonumber
\end{eqnarray}
Therefore $x_{2}=0$. From $G\left( x_{1},x_{2}\right)=0$ and for $x_1\neq0$,
we have
\begin{equation}
x_{1}^{2}+x_{1}+\Delta =0 \, .  \label{QuadPol}
\end{equation}
The roots of this quadratic polynomial are
\begin{equation}
\alpha _{\pm}=-\frac{1}{2}\pm \frac{\sqrt{1-4\Delta }}{2}\, ,  \label{alpha}
\end{equation}
where the symbol $\alpha _{\pm}$ stands for the two different roots of
Eq.~(\ref{QuadPol}). These roots are placed symmetrically with respect
to $x_1=-0.5$. From now on we shall use the symbol $\alpha$ without lower
indices while referring to any of these roots.

According to the Grobman-Hartman theorem \cite{Palis}, the local behavior
near the equilibrium position depends on the linear part (if exists) of the
vector field, in those cases where the eigenvalues of the Jacobian matrix of
the vector field associated to the system have nonzero real part. Hence, in
order to develop local analysis of the equilibrium positions, we calculate

\begin{equation}
J\left( x_{1},x_{2}\right) =\left(
\begin{array}{cc}
0 \quad & \, 1 \\
\frac{1}{2C}\left(-1+\frac{\Delta -x_{2}}{x_{1}^{2}}\right) \quad & \, \frac{%
1}{2C}\left(2C+3+\frac{1}{x_{1}}\right)
\end{array}
\right) .
\end{equation}

Because the equilibrium positions are of type $\left( \alpha ,0\right)$ the
Jacobian matrix in these points is written as
\begin{equation}
J\left( \alpha ,0\right) =\left(
\begin{array}{cc}
0 & 1 \\
g_{1} & g_{2}
\end{array}
\right) ,  \label{eq:jacob}
\end{equation}
where
\begin{eqnarray}
g_{1}&=&\frac{1}{2C}\left( \frac{\Delta }{\alpha ^{2}}-1\right) ,\label{g1}\\
g_{2}&=&\frac{1}{2C}\left( 2C+3+\frac{1}{\alpha }\right) .\label{g2}
\end{eqnarray}
The characteristic polynomial, corresponding to the eingenvalue problem of
Eq.~(\ref{eq:jacob}), is
\begin{equation}
p(\lambda )=\lambda ^{2}-g_{2}\lambda -g_{1} \, , \label{plambda}
\end{equation}
with roots
\begin{equation}
\lambda _{\pm}=\frac{1}{2}g_{2}\pm \frac{1}{2}\sqrt{g_{2}^{2}+4g_{1}}\, .
\label{RLambda}
\end{equation}

We study the possible behavior near the equilibrium positions
depending on the values of the polynomial roots (\ref{RLambda}).
For a complete discussion, definitions and main results, see Refs.
~\cite{Elgoltz,GH,Palis}. In general, one is concerned about whether
the eigenvalues are real or complex, and if their real parts are
greater, equal or less than zero.

The roots (\ref{RLambda}) depend on the value of discriminant
$D=g_{2}^{2}+4g_{1} $; they will be real if $D\geq 0$. Hence, the
following inequality holds:
\begin{equation}
D=1+\frac{1}{C}+\left( \frac{3}{2C}\right) ^{2}+\left( \frac{1}{C}+\frac{3}{%
2C^{2}}\right) \frac{1}{\alpha }+\left( \frac{1}{4C^{2}}+\frac{2\Delta }{C}%
\right) \frac{1}{\alpha ^{2}}\geq 0 .
\end{equation}
This inequality can be written as
\begin{equation}
\left[(2C+3)^2-8C\right]\alpha ^{2}+2(2C+3)\alpha + 8C\Delta + 1 \geq 0 .
\label{complex}
\end{equation}
The polynomial in Eq.~(\ref{complex}) has roots
\begin{equation}
\alpha_{+,-}= \frac{-(2C+3)\pm2\sqrt{2C\left(1-\left[(2C+3)^2-8C\right]%
\Delta\right)}}{(2C+3)^2-8C}.  \label{Croots}
\end{equation}

We shall see soon that the fixed point $(\alpha_{-},0)$ is always a
saddle point. Hence, the current analysis has sense only for
$\alpha_+$. Comparing the expressions for this root given by
Eqs.~(\ref{alpha}) and (\ref{Croots}), solving for $\Delta$ and
substituting the value of $C\approx -0.73$, we find out that roots
(\ref{RLambda}) are real if and only if $\Delta \leq 0.124273$.

\begin{figure}[t]
\centerline{\psfig{file=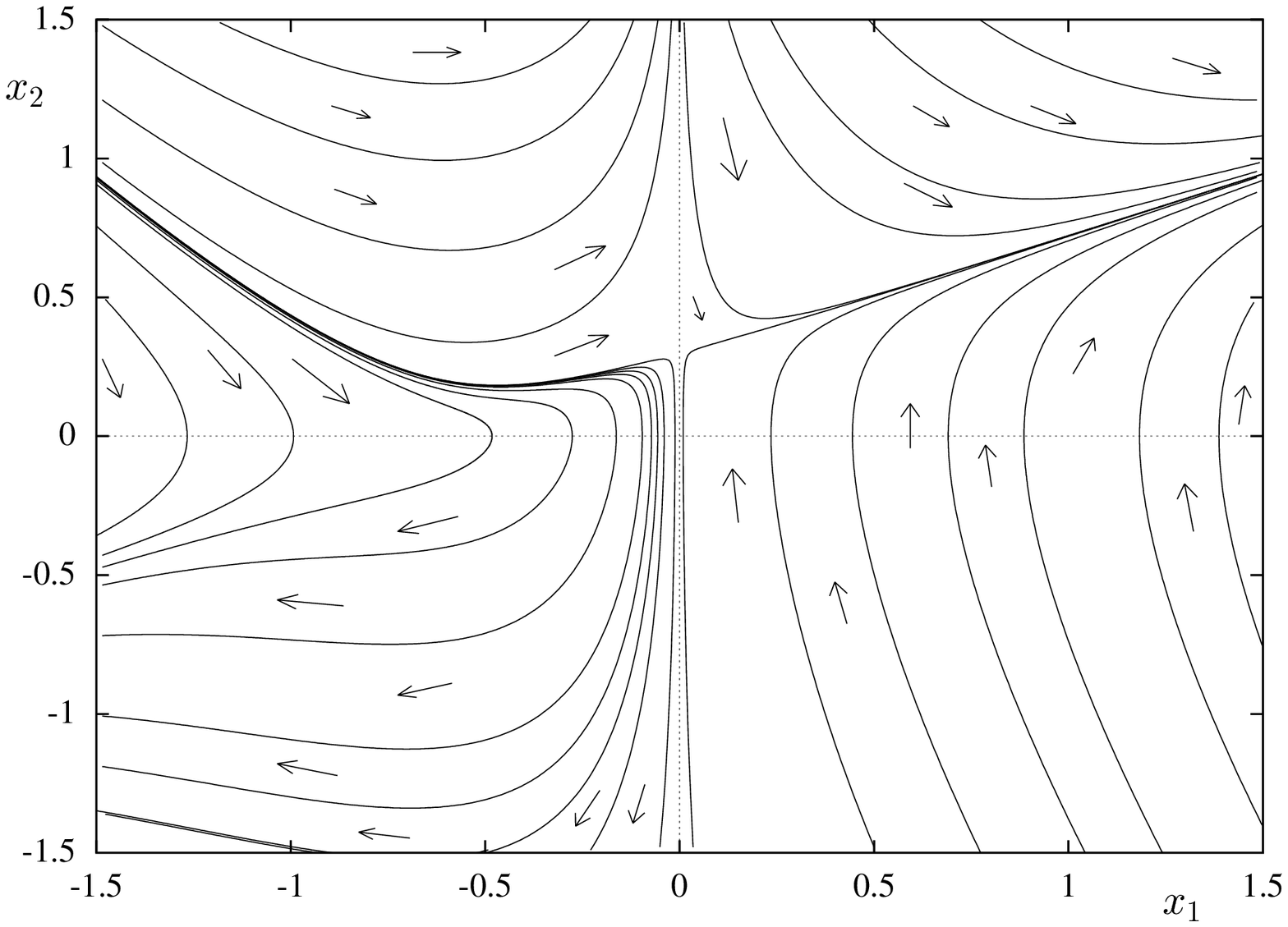,width=10cm} \psfig{file=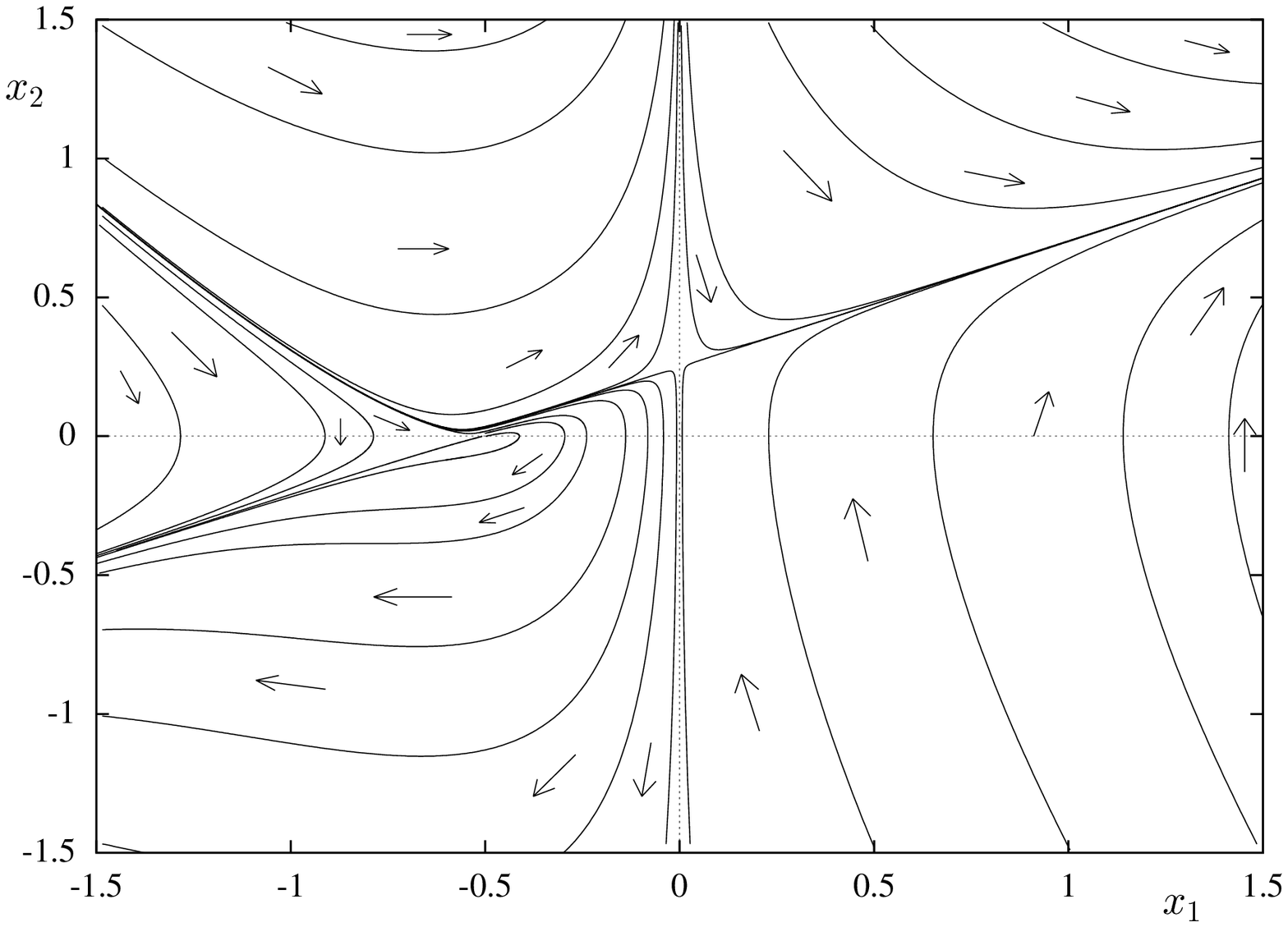,width=10cm}}
\caption{Reduced phase portraits for $\Delta=0.3$, case (a), and for 
$\Delta=0.25$, case (b).}
\label{fig:ppDab}
\end{figure}

Now we are able to classify the fixed points. For $\Delta>0.25$, case (a),
there are not fixed points and the flow is deformed from a laminar one only
near the point $p_s$ (see Fig.~\ref{fig:ppDab}, case (a) and Sec.~
\ref{Point_ps}). Hereafter, in the figures we represent the flow direction by
means of arrows.

For $\Delta=0.25$, case (b), one of the eigenvalues has zero real
part and hence the stability of the unique and degenerated fixed
point $(-0.5,0)$ cannot be determined by linearization. The
remaining eigenvalue has positive real part, then the local
behavior transverse to the center manifold is controlled by the
exponentially expanding flows in the local unstable manifold.
According to Center Manifold theorem for flows \cite{GH}, the
center manifold (it could be not unique) is tangent at $(-0.5,0)$
to the center subspace that, in this case, coincides with the
$x_1$ axis. On the other hand, the unstable manifold is tangent to
the unstable subspace, i.e., the line $x_2=g_2x_1$ with $g_2$, for
this case, being $(2C+1)/2C\simeq 0.315068$ [Fig.~\ref{fig:ppDab},
case (b)]. We recall that each subspace is spanned by the
eigenvectors corresponding to eigenvalues (\ref{RLambda}). Further, 
for $\Delta<0.25$ and $\Delta\neq 0$ we have two
nondegenerate fixed points over the $x_1$ axis symmetrically
distributed with respect to $x_1=-0.5$.

\begin{figure}[h]
\centerline{\psfig{file=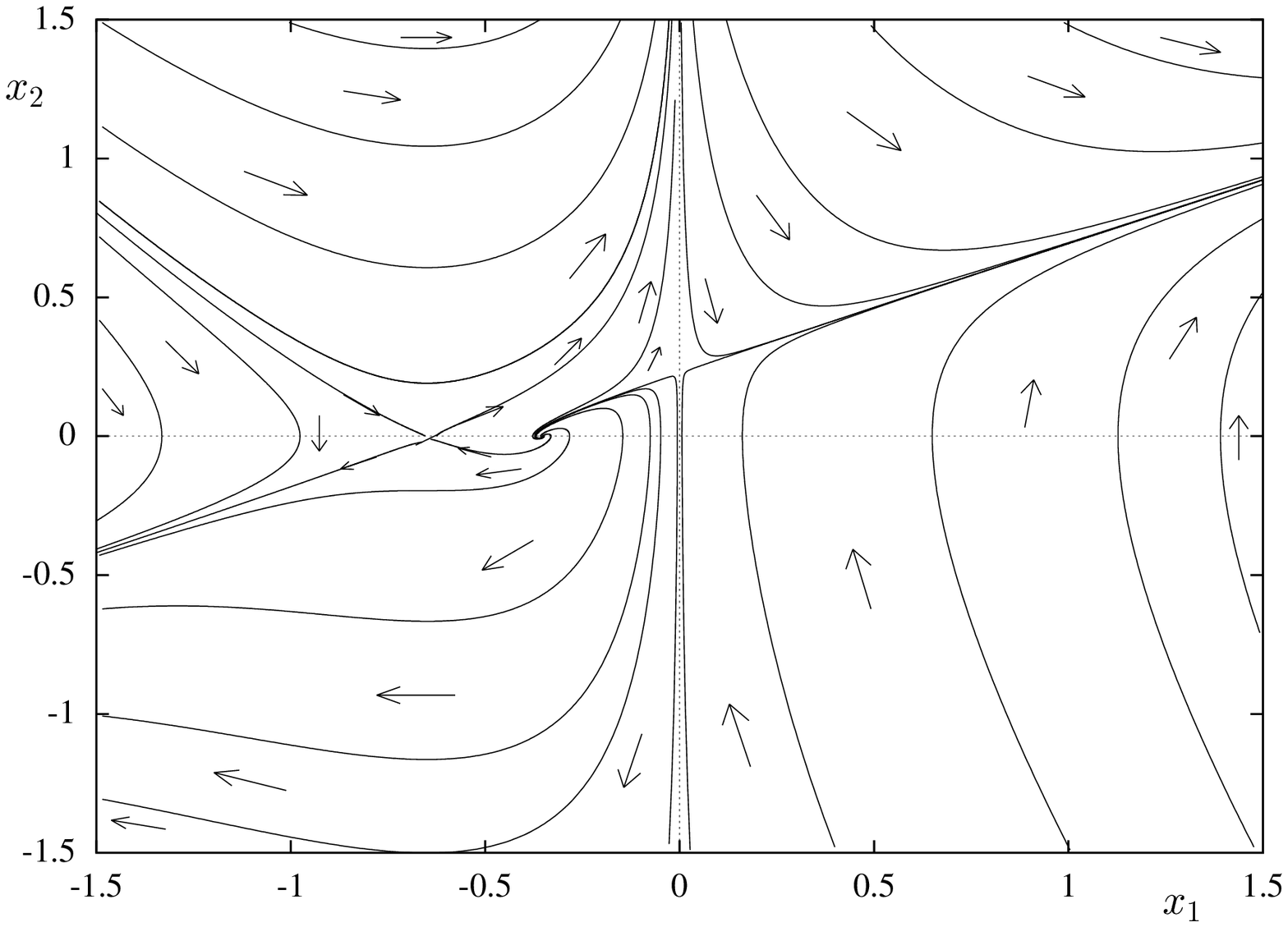,width=10cm} \psfig{file=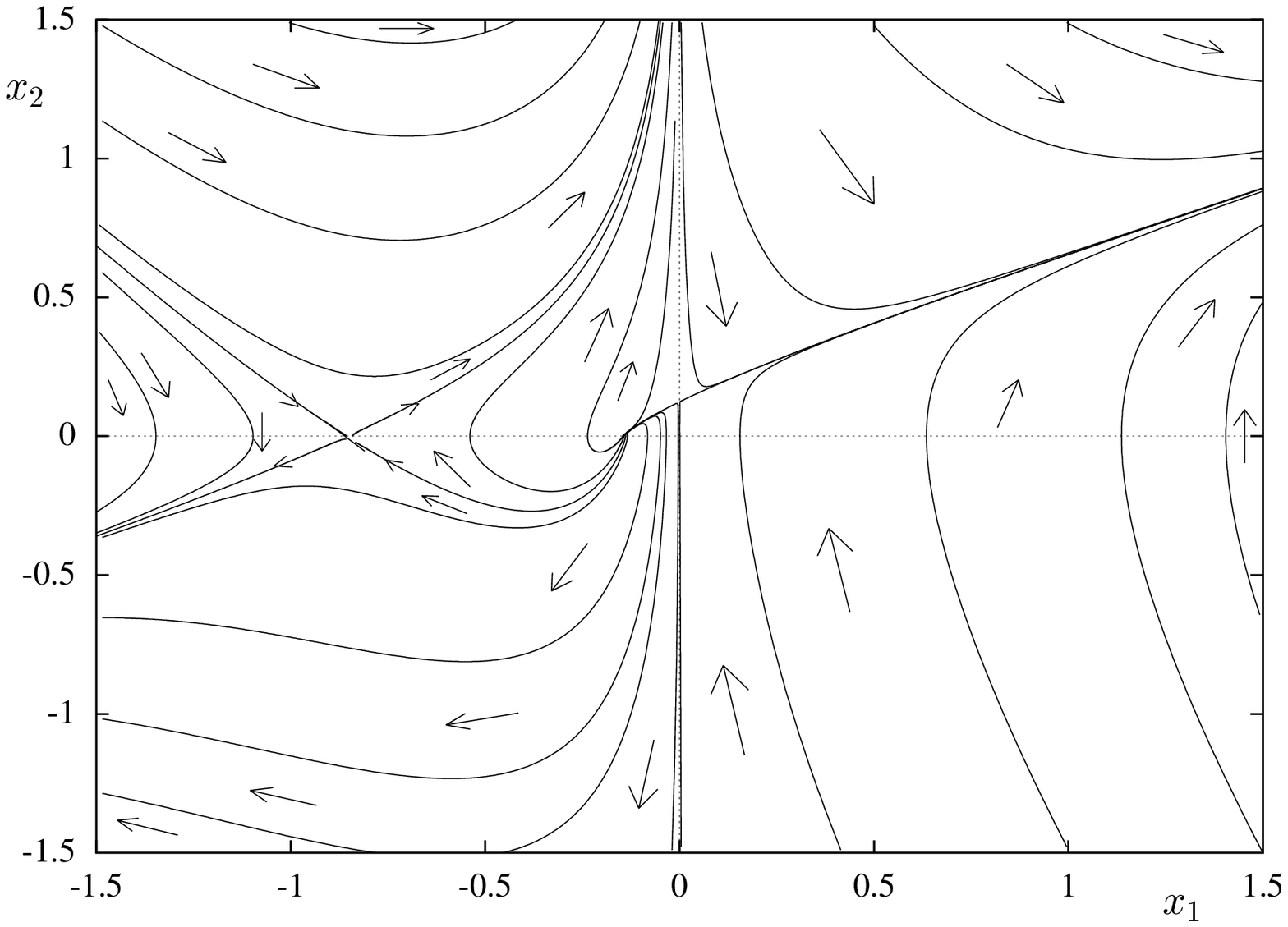,width=10cm}}
\caption{Reduced phase portrait for $\Delta=0.227694$, case (c),
and for $\Delta=0.124273$, case (d).}
\label{fig:ppDcd}
\end{figure}
\begin{figure}[h]
\centerline{\psfig{file=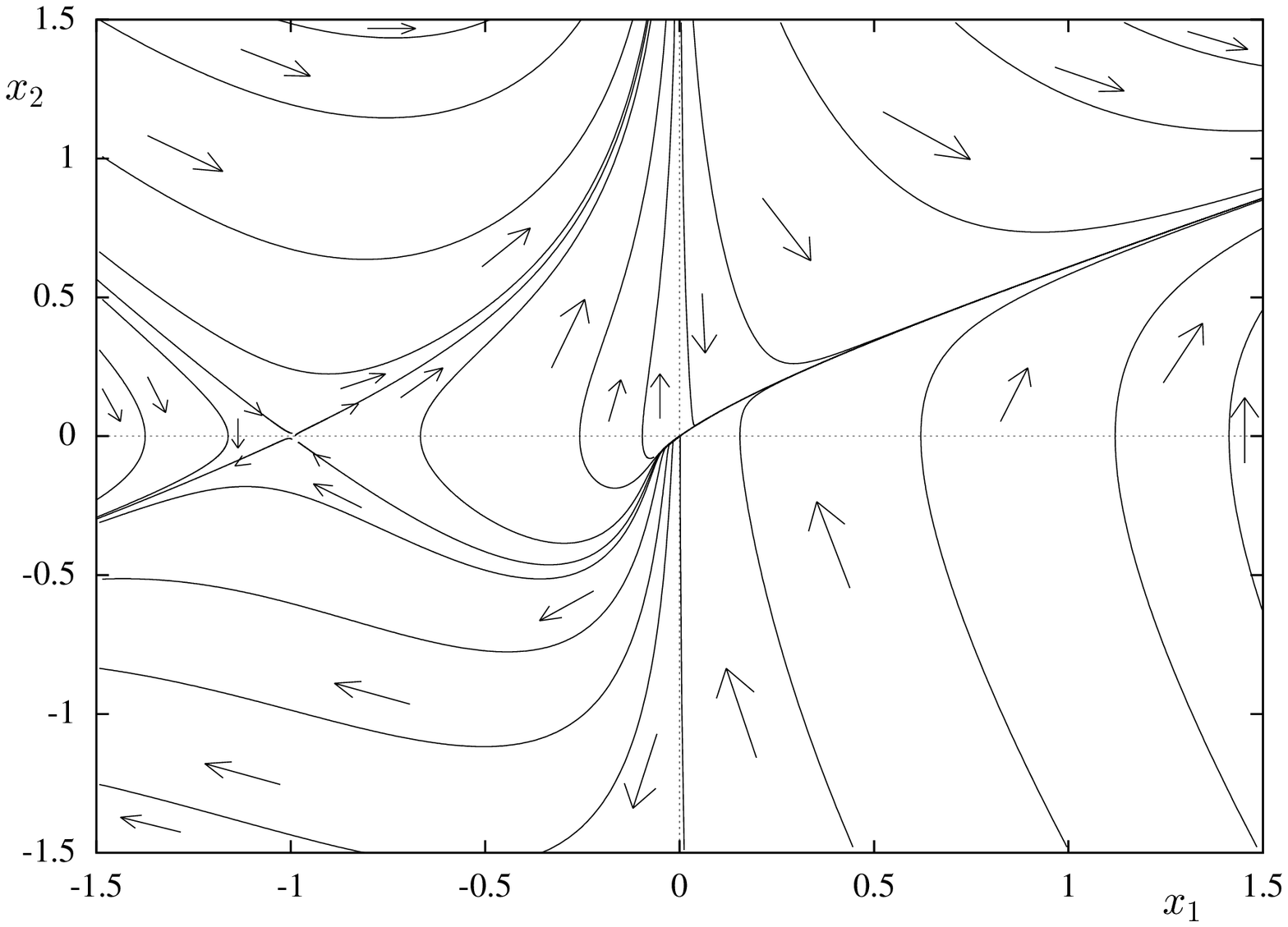,width=10cm} \psfig{file=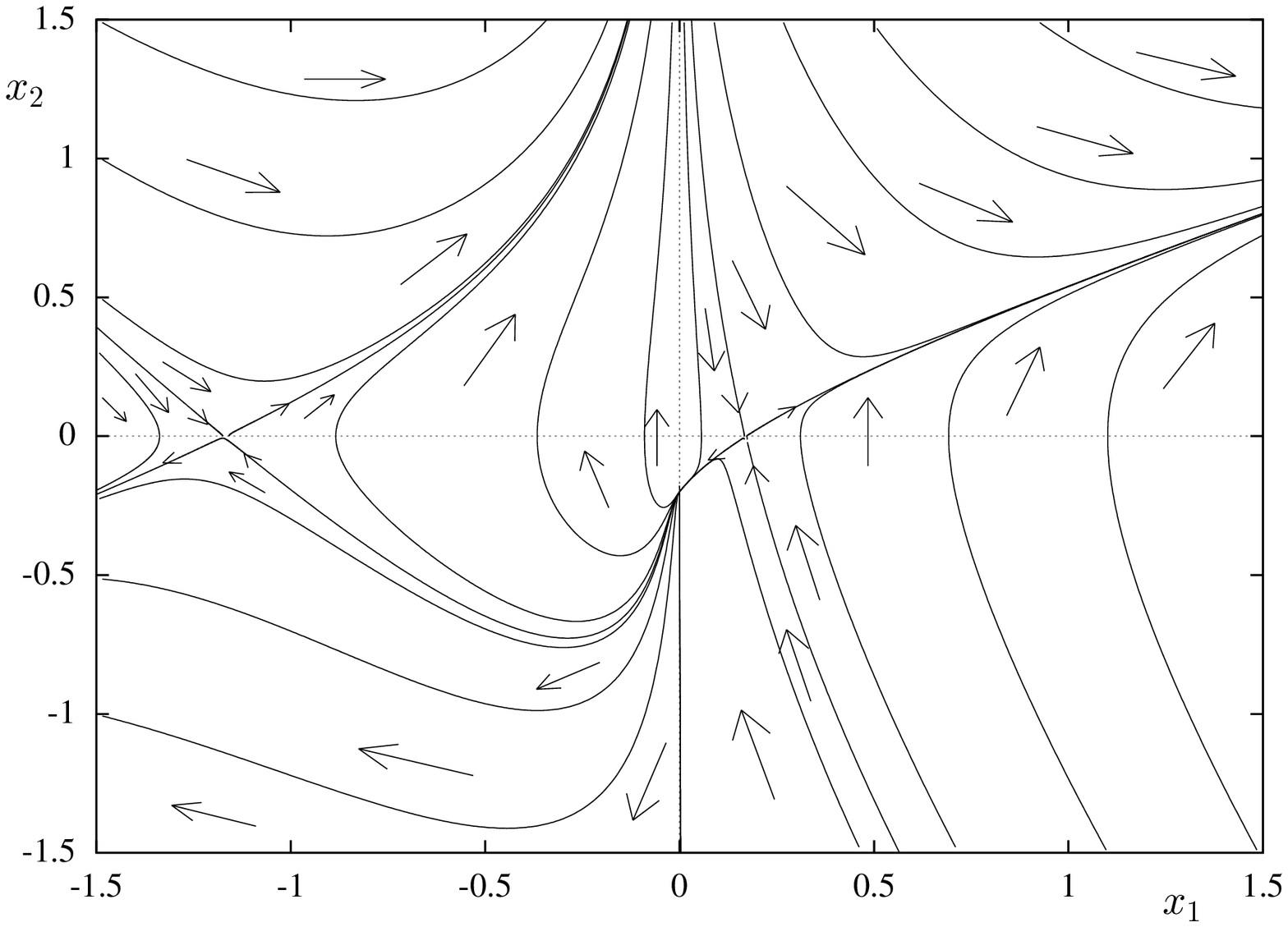,width=10cm}}
\caption{Reduced phase portrait for $\Delta=0$, case (e),
and for $\Delta=-0.2$, case (f).}
\label{fig:ppDef}
\end{figure}

Let us proceed with the fixed points classification paying
attention to the leftmost one. Obviously, the value of $\alpha_-$
for this point is always negative for any $\Delta$ and hence
$g_1>0$. This way, the leftmost fixed point is always a saddle. The
rate $\left|\lambda_+\right|/\left|\lambda_-\right|$ is equal to
$1$ for $\Delta=-0.25+2(C+1)/(2C+3)\simeq 0.227694$, value of
$\Delta$ that corresponds to $\alpha_+=-0.64935$. For $\Delta$
greater (less) than $0.227694$ the trajectories converge to the
saddle point faster (slower) than they diverge from it.

For the rightmost fixed point the situation is richer. First, from
Eq.~(\ref{g2}) one can see that $g_2>0$ for any $\Delta<0.25$ and
$\Delta\neq 0$. For $0<\Delta<0.25$ we have $g_1<0$, then in the
interval $0.124273<\Delta<0.25$, case (c), the rightmost fixed
point is an unstable focus (Fig.~\ref {fig:ppDcd}).
As can be observed in the figure, one of the stable separatrices of
the saddle point seems to be directly spiraling out from the focus.
For $0<\Delta\leq 0.124273$, case (d), the rightmost fixed point
is an unstable node (Fig.~\ref{fig:ppDcd}). In the case $\Delta=0$,
case (e), $g_2<0$ and we have a single saddle located at $(-1,0)$
(see Fig.~\ref{fig:ppDef}). For $\Delta<0$, case (f), the rightmost 
fixed point becomes a saddle (Fig.~\ref{fig:ppDef}) located on the right 
half-plane.

\subsection{Closed orbits}

Note that the sign of the vector field divergence of
system (\ref{DynSys}) does not depend on $\Delta$, thus the value of the
parameter is of no importance while applying the Bendixson
criterion \cite{GH}, which states that no closed path can be found
in plane regions where the vector field divergence is null or not
undergoes sign changes. This criterion, applied to system (\ref{DynSys}),
divides the plane in four regions without divergence sign changes:
\[
\begin{array}{ll}
{\mathcal P}_1 \equiv \{(x_1,x_2); x_1>0, \forall x_2 \},& \qquad
\text{with diverging vector field}, \\
{\mathcal P}_2 \equiv \{(x_1, x_2); -0.64935 < x_1 < 0, \forall x_2 \},& \qquad
\text{with a converging vector field}, \\
{\mathcal P}_3 \equiv \{(-0.64935, x_2); \forall x_2 \},& \qquad
\text{with null divergence, and}\\
{\mathcal P}_4 \equiv \{(x_1,x_2); x_1< -0.64935, \forall x_2 \},& \qquad
\text{with positive divergence}.
\end{array}
\]

\begin{figure}[h]
\centerline{\psfig{file=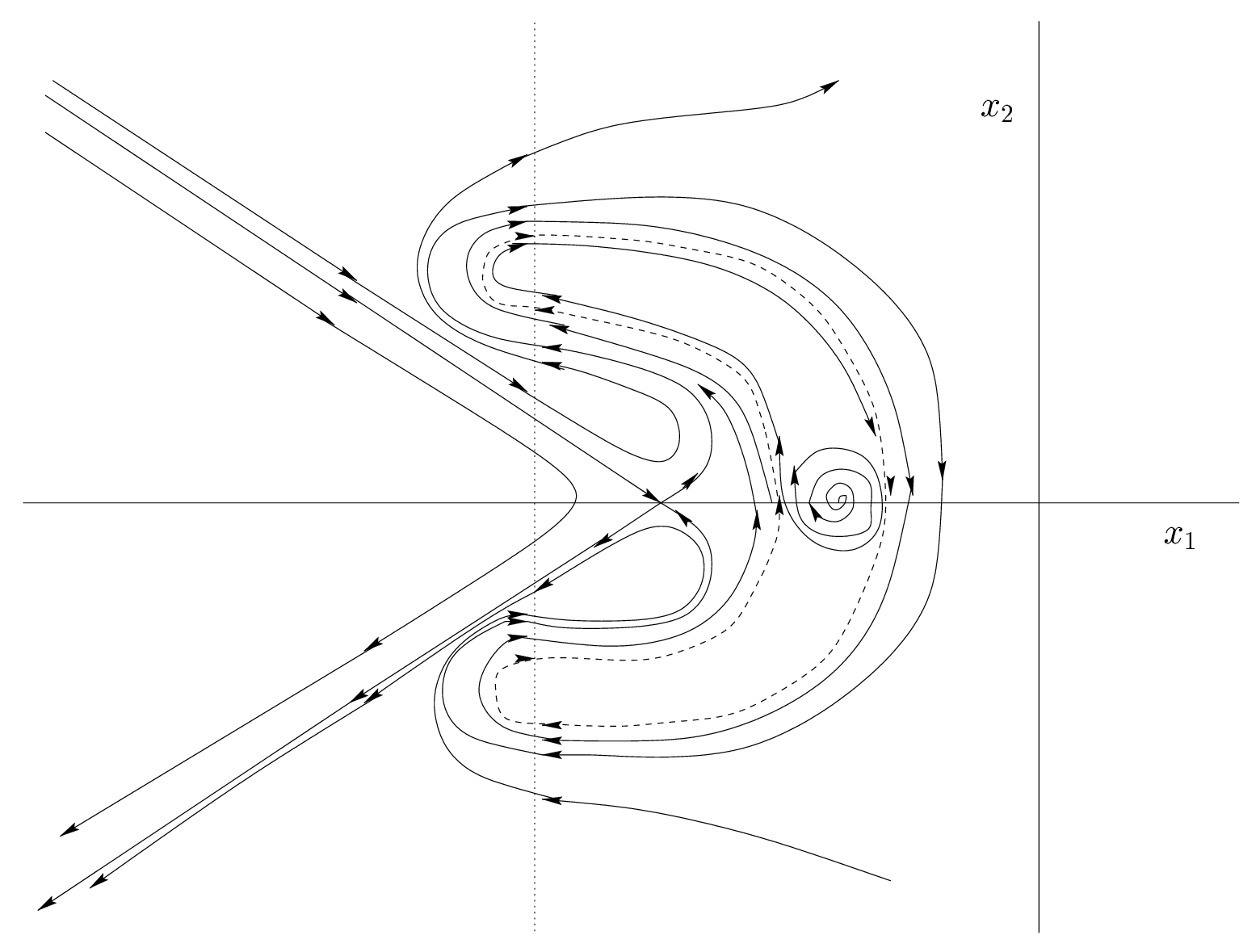,width=10cm}
\psfig{file=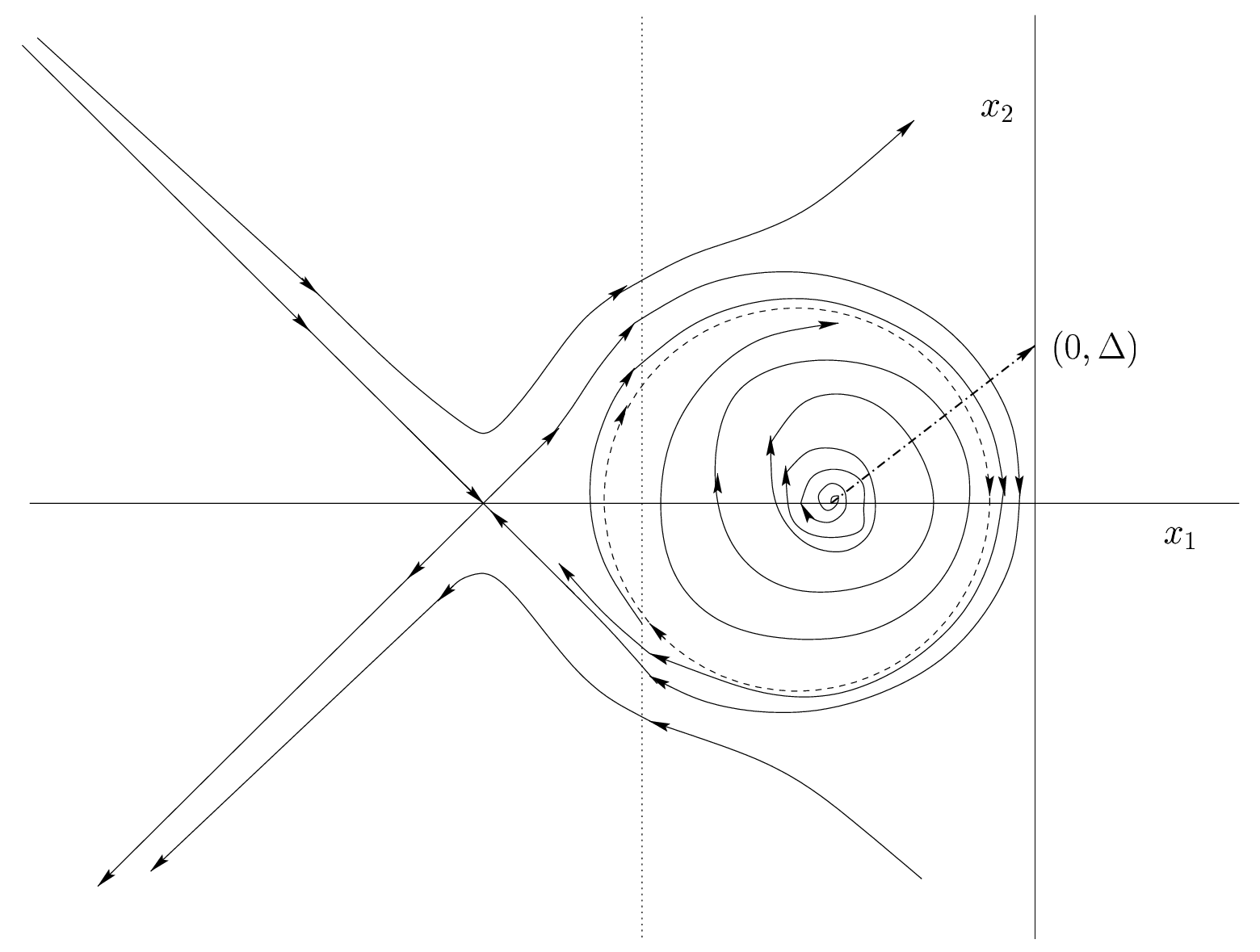,width=10cm}}
\caption{Scheme of a homoclinic orbit with an embedded limit cycle. The
dashed curve represents the limit cycle and the dot-dashed line, the 
trajectory given by Eq.~\ref{Cline}. The dotted line is $x_1=-0.64935$.}
\label{fig:cycles}
\end{figure}

It is worthy to note that, even if both half-planes ($x_1<0,
x_1>0$) are connected through point $p_s$, this connection should
take place in one and only one direction. This way, no closed orbit
could be found in regions ${\mathcal P}_1$ and ${\mathcal P}_3$ or
entirely lying in ${\mathcal P}_2$ or ${\mathcal P}_4 $. Any closed
trajectory should belong to ${\mathcal P}_2\cup{\mathcal
P}_3\cup{\mathcal P}_4$. Then, three possible scenarios with closed
path could take place. First, a homoclinic orbit (i.e., a saddle
connection) enclosing the second fixed point (focus or node, we
will refer to it as $\alpha_+$). The second possibility is a limit
cycle to which converges all of the trajectories starting from
$\alpha_+$ and from which spiral out the orbits. Finally, it could
be possible to find a homoclinic orbit with an embedded limit cycle
around $\alpha_+$. Each of these scenarios has two possible
realizations depending on whether both fixed points lie in
${\mathcal P}_2$ or the saddle lies in ${\mathcal P}_4$ and
$\alpha_+$ lies in ${\mathcal P}_2$. The realizations for the more
general scenario (a homoclinic orbit with an embedded limit cycle)
are schematically represented in Fig.~\ref{fig:cycles}.

One conclude from the left scheme in Fig.~\ref{fig:cycles} that not
closed trajectory will be found with both fixed points lying in
${\mathcal P}_2$ because sign changes in the derivative for the
vector field along $x_1$, at each side of axis $x_2=0$, are not
allowed by the first equation of system (\ref {DynSys}). With regards to
the realization plotted on the right part of Fig.~\ref
{fig:cycles}, we shall see in Sec.~\ref{Point_ps} that for
this range of $x_1$ there is a trajectory (represented by the dot-dashed 
line on the figure) that can be fairly approximated by
\begin{equation}
x_2=\left(1-\frac{x_1}{\alpha_+}\right)\Delta\, ,  \label{Cline}
\end{equation}
then no closed orbit can neither exists here.

\subsection{Flow near point $p_s$}

\label{Point_ps}

As we have already mentioned, a sharp feature of these phase spaces
is the collisions of the trajectories with the border $x_1 = 0$.
Our system possess here a vector field discontinuity that strongly
determines the behavior of the trajectories near this border. Such
a piecewise continuous map is a more general system than those
piecewise differentiable maps previously reported and analyzed in
literature (see, for example, Refs.~\cite{Y1,Y2}).

A particularly interesting situation is exhibited by the
trajectories near the point $p_s=(0, \Delta)$. First of all, let us
note that no statement can be done about the existence or
uniqueness of solutions of system (\ref{DynSys}) at this point. To analyze
the behavior of the flow in the neighborhood of this point we
rewrite the system as
\begin{eqnarray}  \label{dos1}
\hat{x_1} &=& x_2 \, , \\
\hat{x_2} &=& \frac{1}{2C}\left\{(3+2C)\Delta - x_1 - 1 + \left[(3+2C)x_1 +
1\right]\frac{(x_2-\Delta)}{x_1}\right\}.
\end{eqnarray}

We shall look for a solution $x_2 = \Delta + f(x_1)$ for the phase
curves in the neighborhood of the above mentioned point.
Differentiating this expression with respect to $x_1$, taking into
account that $\hat{x_2}=\hat{x_1}{dx_2}/{dx_1}=x_2 {df}/{dx_1}$,
substituting in Eq.~(\ref{dos1}) for the approximation $x_1\ll 1$ and
$f(x_1)\ll 1$ ($x_2\approx \Delta$), we obtain the following
nonhomogeneous linear equation for $f(x_1)$:
\begin{equation}
\frac{df}{dx_1} - \frac{1}{2C\Delta}\frac{f}{x_1} = \frac{(3+2C)\Delta-1}{%
2C\Delta}\, .  \label{fp}
\end{equation}

For $\Delta \neq{1}/{2C}$ the solution of Eq.~(\ref{fp}) is
\begin{equation}
f(x_1) = \left(1-\frac{3\Delta}{1-2C\Delta}\right)x_1 + K
x_1^{{1}/{2C\Delta}}\, ,  \label{f1}
\end{equation}
while for $\Delta={1}/{2C}$,
\begin{equation}
f(x_1) = \frac{3}{2C} \,x_1\ln{x_1} + K x_1 \, ,  \label{f2}
\end{equation}
where $K$ is an integration constant depending on the initial conditions.
Hence, the behavior around the point $p_s$ is described by the following
family of curves
\begin{equation}
x_2= \left\{
\begin{array}{ll}
\label{sp1} \Delta + \left(1-\frac{3\Delta}{1-2C\Delta}\right)x_1 + K x_1^{{1%
}/{2C\Delta}}\, , & \quad \text{if $\Delta \neq {1}/{2C}$}\, , \\
\label{sp2} \frac{1}{2C} + \frac{3}{2C}\,x_1 \ln{x_1} + K x_1 \, , & \quad
\text{if $\Delta = {1}/{2C}$}\, .
\end{array}
\right.
\end{equation}

\begin{figure}[h]
\centerline{\psfig{file=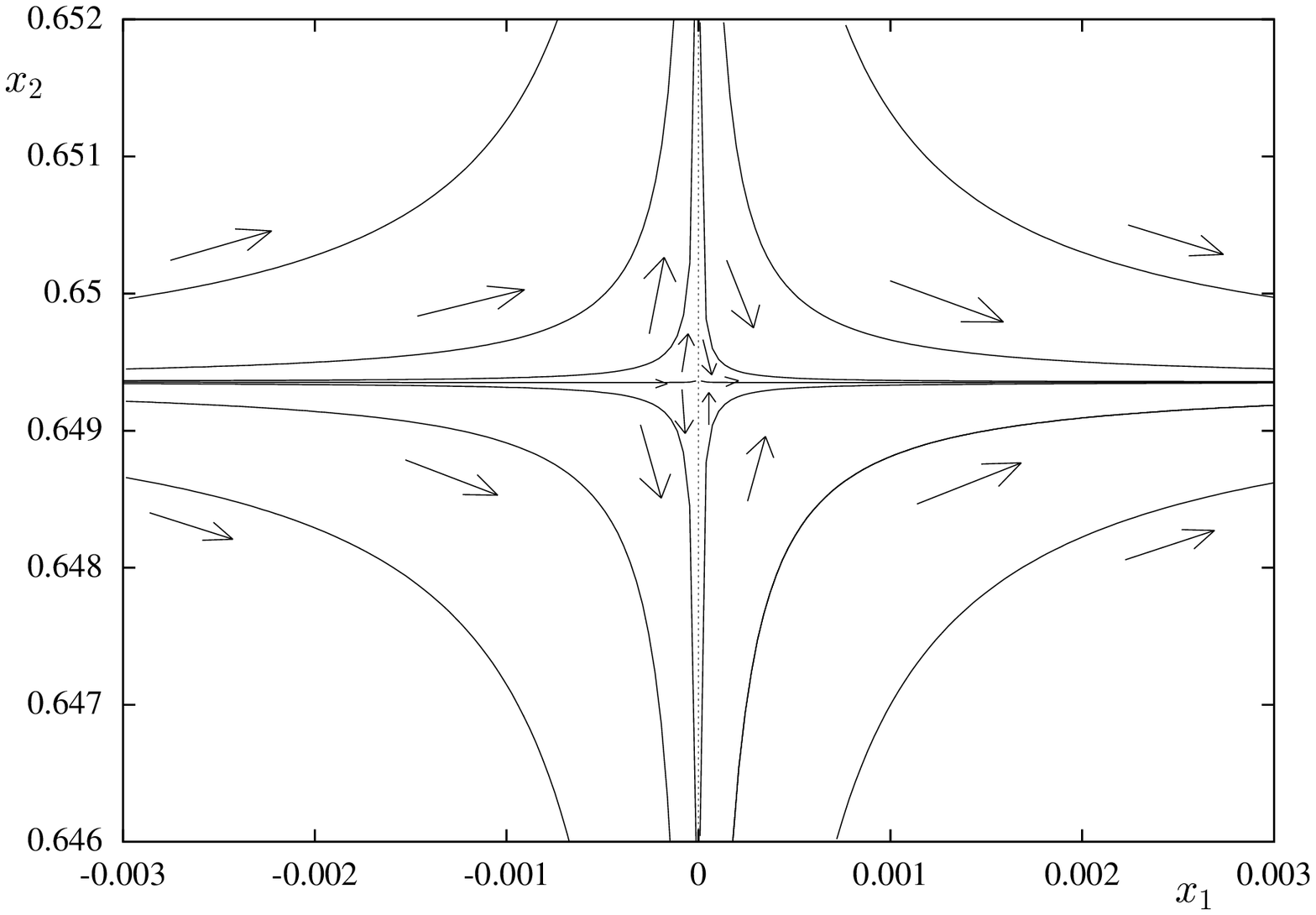,width=10cm} %
\psfig{file=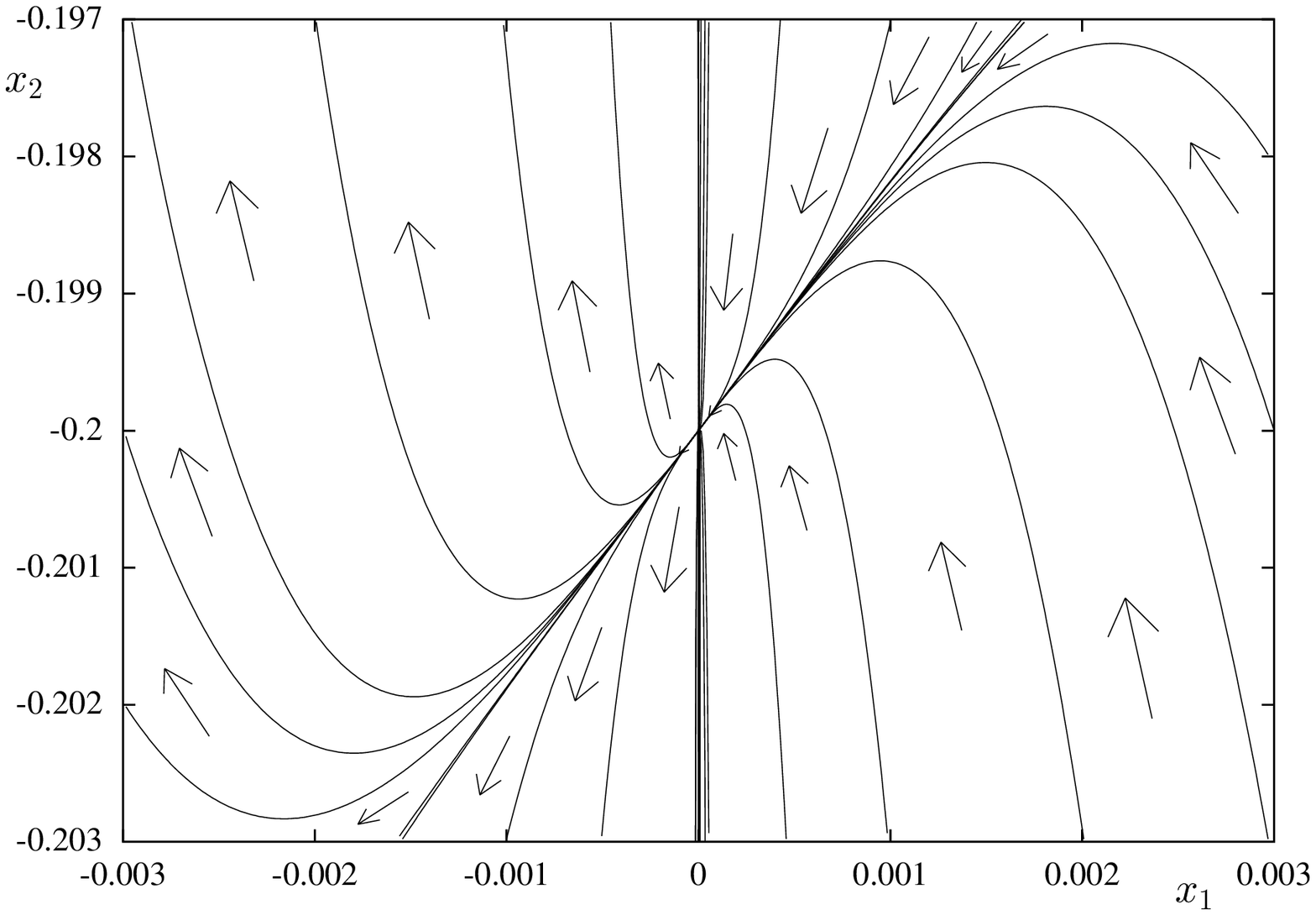,width=10cm}}
\caption{Flow near the point $p_s$ for $\Delta>0$ ($\Delta\simeq0.64935$)
and for $\Delta<0$ ($\Delta\simeq-0.2$).}
\label{fig:psD}
\end{figure}

Concerning the qualitative behavior of the trajectories near the
point $p_s$ there are four interesting intervals of $\Delta$.
First, for $\left|\Delta\right|\gg 1$ the flow seems to ignore the
existence of the special singular point $(0,\Delta)$. All the
trajectories flows to (or from) the $x_2$ axis along parallel
lines with slope $1+3/2C\simeq -1.054794 $. From these lines only
that intercepting the $x_2$ axis at $p_s$ could arrive to or
depart from this point. For positive (but not too large) $\Delta $
the trajectories have the distribution observed in the left part of
Fig.~\ref{fig:psD}.
In this figure we present the case $\Delta=1/(3+2C)\simeq 0.64935$
where the line with $K=0$ coincides with the stationary solution
for $x_1$. Then, for $\Delta>0$ one and only one trajectory could
arrive to singular point $p_s$ and one and only one trajectory
could leave it. It is precisely this trajectory which can be
represented by Eq.~(\ref{Cline}) for $-0.64935<\alpha_+<0 $. For
$\Delta=0$, in the left half-plane, all of the trajectories
between the unstable separatrices of the saddle point and the
ordinates axis move away from $p_s$ diverging from the line
$x_2=x_1$. In the right half-plane, all of the trajectories
converge asymptotically to the solution that behaves as $x_2=x_1$
in the neighborhood of $p_s$ [recall Fig.~\ref{fig:ppDef}, case
(e)]. All of the trajectories converge to $p_s$ or diverge from it
in the same direction. Finally, in the case of negative $\Delta$
(but again with not too large absolute value), the trajectories
converge to $p_s$ or diverge from it tangent to the lines $K=0$
with slope given by Eq.~(\ref{sp1}) (Fig.~\ref{fig:psD}, right part).

\section{The reduced first order equation}

\label{sec:first}

We have already stressed the importance of Eq.~(\ref{MSch2}) for constraining
the possible solutions of Eq.~(\ref{MSch1}). Let us consider now the equation
\begin{equation}
2(C+1)\epsilon \hat{\epsilon }-\epsilon ^{2}-\epsilon -\delta =0 \, ,
\label{FirstOrder}
\end{equation}
with constant $\delta$.

The solutions of this equation also depend on the parameter value.
For $\delta>0.25$, we have
\begin{equation}
\ln \left[ \frac{\left|\epsilon ^{2}+\epsilon +\delta\right| }{B}\right] -%
\frac{1}{\sqrt{\delta -1/4}} \arctan\left( \frac{\epsilon +1/2}{\sqrt{\delta
-1/4}}\right) - \frac{\tau}{C+1}=0\, ,  \label{FO1}
\end{equation}
where $B$ is the integration constant.

For $\delta=0.25$, the solution of Eq.~(\ref{FirstOrder}) is that of the
algebraic equation
\begin{equation}
\left(\epsilon+1/2\right)^{2} - B\exp \left( \frac{\tau}{C+1}
- \frac{1}{\epsilon +1/2}\right)=0\, .  \label{FO2}
\end{equation}
Now, for $\delta<0.25$, and $-0.5-0.5\sqrt{1-4\delta} < \epsilon(\tau) <
-0.5+0.5\sqrt{1-4\delta}$, the solution is obtained from
\begin{equation}
\left|\epsilon^{2}+\epsilon +\delta\right| - B\left(\frac{\sqrt{1-4\delta}%
-2\epsilon-1}{\sqrt{1-4\delta}+2\epsilon+1}\right)^{\frac{1}{\sqrt{1-4\delta}}}
 \exp\left( \frac{\tau}{C+1}\right)=0\,,  \label{FO3}
\end{equation}
and, for $\delta<0.25$ but $\epsilon(\tau) < -0.5-0.5\sqrt{1-4\delta}$ and $%
\epsilon(\tau) > -0.5+0.5\sqrt{1-4\delta}$, the solution results from
\begin{equation}
\left|\epsilon ^{2}+\epsilon +\delta\right| - B\left(\frac{2\epsilon+1
-\sqrt{1-4\delta}}{2\epsilon+1+\sqrt{1-4\delta}}\right)
^{\frac{1}{\sqrt{1-4\delta}}}
\exp\left( \frac{\tau}{C+1}
\right)=0 \,.  \label{FO4}
\end{equation}
Note that the solution for $\delta=0$ is a special case of (\ref{FO4}), i.e.,
\begin{equation}
\epsilon(\tau) = (\epsilon_0+1)\exp\left( \frac{\tau}{2(C+1)}\right)-1\, ,
\label{FO5}
\end{equation}
where $\epsilon_0$ is the initial condition.

\begin{figure}[h]
\centerline{\psfig{file=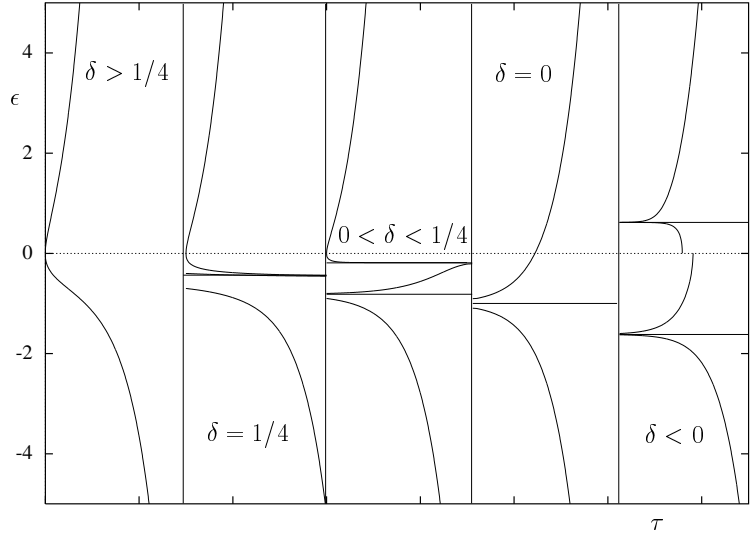,width=10cm}}
\caption{Solutions of the reduced first order equation for different values
of $\delta$. In each of the five regions divided by vertical lines typical
solutions for $\delta$ in the corresponding interval are presented. In every
region each curve branch that goes from left to right is a solution with
different initial values.}
\label{fig:delta}
\end{figure}

Possible behaviors of solutions of Eq.~(\ref{FirstOrder}) for
different values of $\delta$ are summarized in Fig.~\ref
{fig:delta}.
The five interesting intervals are represented by vertical bands in
the above plot. In each band the curves branch spreading from left
to right is a solution starting from a different initial condition.
We can see that all of the solutions are monotonic. In some cases
they evolve unbounded, in some other cases bounded by the
stationary solutions given by the roots of
\[
\epsilon ^{2}+\epsilon+\delta =0\, .
\]

\section{Conclusions}

\label{sec:conclu}

We introduced the Stewart-Lyth inverse problem as the
determination of the inflaton potential through expressions for the
first slow-roll parameter obtained as a solution of differential
equations. These equations were derived from the Stewart-Lyth equations
for the spectral
indices. We tested the feasibility of the method by solving the
problem with constant spectral indices as input, corresponding with
power-law inflation.

To draw conclusions about the behavior of general
$\tau$-depending solutions we analyze reduced equations using the
spectral indices as parameters. The phase space for the reduced
second order equation is richer than the generic ones, particularly
due to the singularity $\epsilon=0$ and to the existence of the
special singular point $[0, 0.5(n_s-1)]$ near which the flow is
deformed even when there are not fixed points. Do not exist
cyclical orbits given by the reduced second order equation for any
value of $n_s$. The condition for the existence of stationary
solutions with positive first slow-roll parameter is $n_s<1$, and
there is only one such a solution for every constant value of the
scalar spectral index.

The reduced first order equation was solved for any value of the
tensorial spectral index. Five possible behaviors were found for
the solutions: stationary, asymptotically increasing,
asymptotically decreasing, boundlessly increasing, and boundlessly
decreasing. The singularity $\epsilon=0$ was also observed, except
in the case $n_T=0$. The condition for the existence of stationary
solutions with positive first slow-roll parameter is $n_T<0$, and
there is only one such a solution for every constant value of the
tensorial spectral index.

In general, for solutions of the Stewart-Lyth inverse problem with
smoothly and slowly changing spectral indices in the expected range
of values, we shall find that the trajectories would be confined in one of
the sectors of the extended phase space divided by
the axis $\epsilon=0$. The exception could be those system with the
tensorial index crossing through value $n_T=0$ while the scalar
index is greater than $1$. Due to the lack of periodic solutions
for the reduced equations, periodic, quasiperiodic, or chaotic
extended solutions should not be expected. Suitable (and unique in
each case) power-law solutions will exist if and only if $n_s<1$
and (consistently with the definition of power-law inflation)
$n_T<0$. Taking this last result into account, estimations of the
scalar index based in the recent
observations by the collaborations Boomerang and Maxima-1
\cite{observ,observ2} indicate that, despite its simplicity, power-law
scenario is still a good
candidate for the inflationary stage of the early universe.

A great inside about general properties of scale-dependent
solutions for smoothly and slowly changing spectral indices was
gained in this work. Further efforts will be focus on obtaining
solutions for the Stewart-Lyth inverse problem, i.e., explicit
functional forms of the inflationary potential.

\acknowledgments

This research was supported in part by the CONACyT grant No. 32138-E and the
Sistema Nacional de Investigadores (SNI). The work of one of the authors (RM) 
was partially supported by project DGAPA IN122498. We want to thank 
Andrew Liddle and Eckehard Mielke for helpful discussions.

\end{document}